\documentclass[onecolumn,a4paper]{IEEEtran}

\usepackage{tikz}
\usepackage{epsfig}
\usepackage{epstopdf}
\usepackage{siunitx}
\usepackage{pbox}
\usepackage{makecell}
\usepackage{multirow}
\usepackage{amssymb}
\usepackage{amsmath}
\usepackage[normalem]{ulem}

\usepackage{blindtext}
\usepackage{etoolbox}
 
\makeatletter

\docsvlist{\@oddhead,\@evenhead,\ps@headings,\ps@IEEEtitlepagestyle,\ps@IEEEpeerreviewcoverpagestyle}
\makeatother

\begin{document}

\title{Accuracy of \emph{Y}-function methods for parameters extraction of two-dimensional FETs across different technologies}

\author{Anibal Pacheco-Sanchez, David Jim\'enez 

\thanks{This work has received funding from the European Union’s Horizon 2020 research and innovation programme under grant agreements No GrapheneCore2 785219 and No GrapheneCore3 881603, from Ministerio de Ciencia, Innovación y Universidades under grant agreement RTI 2018-097876-BC21(MCIU/AEI/FEDER, UE). This  article  has been partially  funded  by  the  European  Regional  Development  Funds  (ERDF)  allocated  to  the  Programa Operatiu FEDER de Catalunya 2014-2020, with the support of the Secretaria d’Universitats i Recerca of the Departament d’Empresa i Coneixement of the Generalitat de Catalunya for emerging technology clusters to  carry  out  valorization  and  transfer  of  research  results.  Reference  of  the  GraphCAT  project:  001-P-001702. \newline \indent A. Pacheco-Sanchez and D. Jim\'enez are with the Departament d'Enginyeria Electr\`{o}nica, Escola d'Enginyeria, Universitat Aut\`{o}noma de Barcelona, Bellaterra 08193, Spain, e-mails: AnibalUriel.Pacheco@uab.cat, david.jimenez@uab.cat}
}
\maketitle
\makeatletter
\def\ps@IEEEtitlepagestyle{
  \def\@oddfoot{\mycopyrightnotice}
  \def\@evenfoot{}
}
\def\mycopyrightnotice{
  {\footnotesize
  \begin{minipage}{\textwidth}
  \centering
  This paper is a postprint of a paper submitted to and accepted for publication in Electronics Letters and is subject to Institution of Engineering and Technology Copyright. The copy of record is available at the IET Digital Library
  \end{minipage}
  }
}
\begin{abstract}
\boldmath
The accuracy of contact resistance values of two-dimensional field-effect transistors extracted with the \textit{Y}-function considering the impact of the intrinsic mobility degradation is evaluated here. The difference between methodologies that take this factor into account and ignore it is pointed out by a detailed analysis of the approximations of the transport model used for each extraction. In contrast to the oftenly used approach where the intrinsic mobility degradation is neglected, a \textit{Y}-function-based method considering a more complete transport model yields contact resistance values similar to reference values obtained by other intricate approaches. The latter values are more suitable also to describe experimental data of two dimensional devices of different technologies. The intrinsic mobility degradation factor of two-dimensional transistors is experimentally characterized for the first time and its impact on the device performance is described and evaluated.
\end{abstract}
%
%

\IEEEpeerreviewmaketitle

\section{Introduction}

The contact resistance $R_{\rm{C}}$ of a field-effect transistor (FET) can be extracted from a relation of a transistor's drift-diffusion (DD) drain current $I_{\rm{D}}$ at the linear regime to its transconductance $g_{\rm{m}}$. This relation is generally known as the $Y$-function such as $Y=I_{\rm{D}}g_{\rm{m}}^{-0.5}$ \cite{Ghi88}. In two-dimensional (2D) emerging transistor technologies where a DD approach can still be considered at the linear operation regime \cite{Anc10}, \cite{MarBad18}, different $Y$-function based methodologies (YFMs) for the $R_{\rm{C}}$ extraction have been proposed \cite{ChaYan13}-\cite{PacFei20}. However, these methodologies consider different $I_{\rm{D}}$ model approximations, i.e., different $R_{\rm{C}}$ values can be found for a device under similar bias conditions. A comparison of such results with different YFMs for 2D FETs is reported here for the first time according to the authors' knowledge.

YFMs are an immediate alternative to test-structure based extraction approaches which robustness is still under discussion for 2D FETs, e.g., the transfer length method (TLM) \cite{ChaZhu14}, \cite{DirVen20}. An evaluation of the YFM extracted values accuracy is to evaluate the closeness of the considered $I_{\rm{D}}$ model including the extracted parameters \cite{ChaZhu14}, \cite{PacFei20}, \cite{PacCla16} with the experimental data. This verification step has been either rarely provided in 2D FET technologies \cite{NaShi14}-\cite{UrbLup20} or it has been considered for one single transfer curve at low fields of a specific technology \cite{ChaZhu14}. Notice that providing a comparison for the curve of the $Y$-function \cite{ChaYan13}, \cite{ChaZhu14}, \cite{NaShi14}, is not a complete validation since the impact of all the parameters involved in the $I_{\rm{D}}$ model is not included, e.g., mobility degradation effects \cite{Ghi88}. Hence, the validness of the extracted values, including $R_{\rm{C}}$, under certain approximations of $I_{\rm{D}}$ is not clear enough. In this work, $R_{\rm{C}}$ of different emerging transistor technologies with 2D materials as channels, namely, molybdenium disulfide (MoS$_2$), black-phosphorus (BP), tungsten diselenide (WSe$_2$) and graphene (G) have been extracted by means of YFMs relying on different approximations of DD $I_{\rm{D}}$ models. The implications of each approach is discussed in detail by analyzing other extracted values generally neglected. 

\section{Y-function-based-methods for 2D FETs} \label{ch:YFMs}
In general, the internal electron drain current can be described within the DD approach at the linear regime as \cite{Sch06}

\begin{equation} 
I_{\rm{D}} = k \left(V_{\rm{GS,i}} - V_{\rm{th}} - \SI{0.5}{}{V_{\rm{DS,i}}} \right) V_{\rm{DS,i}},
\label{eq:YFM_Id_intr_app}
\end{equation} 

\noindent where $k=\mu_{\rm{eff}} C_{\rm{ox}} w_{\rm{g}}/L$ with $\mu_{\rm{eff}}$ as the effective mobility, $C_{\rm{ox}}$ the oxide capacitance, $w_{\rm{g}}$ the gate capacitance and $L$ the channel length, $V_{\rm{th}}$ is the charge threshold voltage \cite{Ghi88} and $V_{\rm{GS,i/DS,i}}$ are the intrinsic gate-to-source and drain-to-source voltages. The latter are considered as $
V_{\rm{GS,i}} \approx V_{\rm{GS}} - I_{\rm{D}}R_{\rm{C}}/{2}$ and $
V_{\rm{DS,i}} \approx V_{\rm{DS}} - I_{\rm{D}}R_{\rm{C}}$, respectively, with $R_{\rm{C}}$ embracing the contribution of the source/drain contact resistances $R_{\rm{C,S/C,D}}$ and $V_{\rm{GS/DS}}$ as the external gate-to-source/drain-to-source voltage. In terms of the more practical external voltages and using $\mu_{\rm{eff}}=\mu_{0}/\left[1+\theta_{\rm{ch}}\left(V_{\rm{GS}}-V_{\rm{th}}-\Delta\right)\right]$ \cite{MerBor72} with $\theta_{\rm{ch}}$ as the intrinsic mobility degradation coefficient due to vertical fields, the drain current equation is given by

\begin{equation} 
I_{\rm{D}} \approx \beta \frac{\left(V_{\rm{GS}} - V_{\rm{th}} - \Delta \right)}{1+\theta \left(V_{\rm{GS}} - V_{\rm{th}} - \Delta \right) } V_{\rm{DS}},
\label{eq:Id_yfm}
\end{equation}

\noindent and the total resistance $R_{\rm{tot}}$ can be expressed as

\begin{equation}
R_{\rm{tot}} = \frac{V_{\rm{DS}}}{I_{\rm{D}}} \approx \left[\frac{1}{\beta\left(V_{\rm{GS}}-V_{\rm{th}}-\Delta\right)} + \frac{\theta_{\rm{ch}}}{\beta}\right] + \frac{\theta_{\rm{C}}}{\beta} \equiv R_{\rm{ch}} + R_{\rm{C}},
\label{eq:rch}
\end{equation}

\noindent where $\beta=\mu_{\rm{0}} C_{\rm{ox}} w_{\rm{g}}/L$ with the low field mobility $\mu_{0}$ and $\theta = \theta_{\rm{ch}} + \theta_{\rm{C}}$ is the extrinsic mobility degradation coefficient \cite{Ghi88}, \cite{HaoCab85} with $\theta_{\rm{C}}=\beta R_{\rm{C}}$. For the special case of GFETs, Eq. (\ref{eq:Id_yfm}) can be modified in a way that the charge control is influenced by the Dirac voltage $V_{\rm{Dirac}}=V_{\rm{GS}}\vert_{\rm{min}(\mathit{I}_{\rm{D}})}$ rather than by a $V_{\rm{th}}$ \cite{PacFei20}.  

In contrast to other approaches in 2D FETs where the potential is considered constant along the channel at low fields \cite{ChaZhu14}, \cite{NaShi14}, in this work the charge control relation considers the contribution of $\Delta=\SI{0.5}{} V_{\rm{DS}}$ as an average of the inhomogenous potential along the 2D channel \cite{MarBad18}, \cite{ZhuPer09}, \cite{WuLi16}. This condition embraces more realistic and practical bias scenarios, e.g., high $\vert V_{\rm{DS}} \vert$. More importantly, most of 2D FET studies neglect the impact of $\theta_{\rm{ch}}$ in Eq. (\ref{eq:Id_yfm}) \cite{ChaZhu14}-\cite{ZhuPar16} without providing a solid argumentation for it but just claiming a challenging experimental characterization of such parameter \cite{ChaYan13}. This is the fundamental difference with the YFM introduced here for 2D FETs. In this work two different YFMs, one without and another with the effect of $\theta_{\rm{ch}}$, namely YFM$_{1}$ and YFM$_{2}$, have been analyzed based on approximations considered for Eq. (\ref{eq:Id_yfm}). 


For YFM$_1$, $\theta_{\rm{ch}}$ has been neglected in Eq. (\ref{eq:Id_yfm}) which corresponds to the most common $Y$-function based extraction methodology applied to 2D FETs \cite{ChaYan13}-\cite{UrbLup20}. If $C_{\rm{ox}}$ is known for the device under study, the methodology described in \cite{ChaZhu14} can be followed. Otherwise, the extraction of a contact resistance $R_{\rm{C,1}}$ must be determined as the difference between the device total resistance $R_{\rm{tot}}$ and a channel resistance $R_{\rm{ch,1}}(\approx 1/\left[\beta\left(V_{\rm{GS}}-V_{\rm{th}}-\Delta\right)\right]$, see Eq. (\ref{eq:rch}) $)$ obtained from the reduced form of Eq. (\ref{eq:Id_yfm}), i.e. $R_{\rm{C,1}}=R_{\rm{tot}}-R_{\rm{ch,1}}$. Notice that $\Delta$ is usually ommitted as well in this extraction approach \cite{ChaYan13}-\cite{UrbLup20}, however, for a fair comparison it has been considered here. $\beta$ has been obtained from the slope of the corresponding $Y$-function.  

In YFM$_2$, $\theta_{\rm{ch}}$ has been considered in Eq. (\ref{eq:Id_yfm}) for the extraction of $R_{\rm{C,2}}$. $V_{\rm{th}}$ and $\beta$ are obtained from the intercept and maximum slope, respectively, of the $Y$-function considering Eq. (\ref{eq:Id_yfm}), i.e., $Y = \sqrt{\beta V_{\rm{DS}}} \left(V_{\rm{GS}}-V_{\rm{th}} - \Delta \right)$, plotted over $V_{\rm{GS}}$. The maximum point of the derivative with respect to $V_{\rm{GS}}$ of a $Y$-function normalized to $I_{\rm{D}}$, i.e., $Y/I_{\rm{D}}=[1+\theta(V_{\rm{GS}}-V_{\rm{th}}-\Delta)]/\sqrt{\beta V_{\rm{DS}}}$, yields a value for $\theta$. Values for $\theta_{\rm{ch}}$ and $R_{\rm{C,2}}$ are obtained from the intercept and slope, respectively, of $\theta$ as a function of $\beta$ built for at least two different $V_{\rm{DS}}$. Notice that the channel resistance obtained with YFM$_2$, i.e., with Eq. (\ref{eq:Id_yfm}) considering $\theta_{\rm{ch}}$, is given by $R_{\rm{ch,2}}\approx \left[1+\theta_{\rm{ch}}\left( V_{\rm{GS}}-V_{\rm{th}}-\Delta \right)\right]/\left[\beta\left(V_{\rm{GS}}-V_{\rm{th}}-\Delta\right)\right]$ (see Eq. (\ref{eq:rch})), however, $R_{\rm{C,2}}$ is extracted in YFM$_2$ without using $R_{\rm{ch,2}}$. This approach, previously introduced for silicon-based devices \cite{KarChe14}, has been adapted and succesfully applied to other emerging transistor technologies \cite{PacCla16} and use for the first time here for 2D FETs. 

As pointed out by \cite{ChaZhu14}, YFMs can mislead a precise estimation of the mobility due to a gate-voltage dependence of the extracted $R_{\rm{C}}$. However, the mobility estimation is out of the scope of this work. Furthermore, the $R_{\rm{C,2}}$ extracted with YFM$_2$ has been found to be in close agreement with bias-dependent contact resistance values of other emerging Schottky-type devices \cite{PacCla20}. 

The above methodologies are also applicable to $p$-type transistors by considering hole transport in Eqs. (\ref{eq:YFM_Id_intr_app}) and (\ref{eq:Id_yfm}).

\section{$R_{\rm{C}}$ of 2D FETs with different approaches}

The YFMs discussed above have been applied to fabricated 2D FET technologies \cite{SanPar17}-\cite{HsuWan11} with different device footprints and different 2D channel materials. Extracted parameters with the YFMs for each device is listed in Table \ref{tab:rc_exp} as well as some device dimensions. $\theta_{\rm{ch}}$ has been extracted with YFM$_2$ only since YFM$_1$ neglects its impact. 


\begin{table} [!htb] 
\begin{center}
\caption{Device dimensions and extracted parameters of 2D FETs.}
\begin{tabular}{c|c||c|c|c|c}

ref. & \makecell{$w_{\rm{g}}/L_{\rm{g}}$\\ (\SI{}{\micro\meter}/\SI{}{\nano\meter})}  & \makecell{$R_{\rm{C1}}\cdot w_{\rm{g}}$\\ $(\SI{}{\kilo\ohm}\cdot\SI{}{\micro\meter})$} & \makecell{$R_{\rm{C2}}\cdot w_{\rm{g}}$\\ $(\SI{}{\kilo\ohm}\cdot\SI{}{\micro\meter})$} & \makecell{$\theta_{\rm{ch}}$\\ $(\SI{}{\volt^{-1}})$} & \makecell{$\theta$\\ $(\SI{}{\volt^{-1}})$}\\\hline \hline

\multicolumn{6}{c}{MoS$_2$FETs} \\ \hline

\cite{SanPar17} & $\SI{10}{}/\SI{150}{}$ &  \SI{13.2}{} & \SI{9.8}{} & \SI{0.32}{} & \SI{1.37}{} \\ \hline
\cite{SanGho15} & $\SI{20}{}/\SI{250}{}$ &  \SI{15.4}{} & \SI{9.9}{} & \SI{1.15}{} & \SI{3.71}{} \\ \hline
\cite{IllBan19} & $\SI{0.8}{}/\SI{400}{}$ & \SI{80.1}{} & \SI{39.8}{} & \SI{0.01}{} & \SI{0.04}{} \\ \hline
\cite{ChaYan13} & $\SI{3}{}/\SI{1000}{}$  &\SI{143}{} & \SI{106}{} & \SI{2.31}{} & \SI{7.07}{}\\ \hline

\multicolumn{6}{c}{BPFETs} \\ \hline

\cite{LiYu19} & $\SI{2.3}{}/\SI{100}{}$ &\SI{1.03}{} & \SI{0.35}{} & \SI{1.58}{} & \SI{2.36}{} \\ \hline
\cite{YanCha17} & $\SI{10}{}/\SI{200}{}$  &\SI{16.8}{} & \SI{8.95}{} & \SI{0.96}{} & \SI{1.21}{} \\ \hline
\cite{YarHar19} & $\SI{3.16}{}/\SI{300}{}$  & \SI{1.8}{} & \SI{1.4}{} & \SI{0.03}{} & \SI{0.39}{} \\ \hline

\multicolumn{6}{c}{WeS$_2$FETs} \\ \hline

\cite{HeiPal20} & $\SI{8}{}/\SI{4000}{}$ &\SI{310}{} k & \SI{109}{} k & \SI{0.06}{} & \SI{0.48}{} \\ \hline
\cite{FanChu12} & $\SI{1}{}/\SI{9400}{}$  &\SI{73}{} & \SI{69}{} & \SI{1.41}{} & \SI{5.13}{} \\ \hline

\multicolumn{6}{c}{GFETs} \\ \hline

\cite{WuZou16} & $\SI{20}{}/\SI{60}{}$ &\SI{0.22} & \SI{0.19}{} & \SI{0.20}{} & \SI{3.18}{} \\ \hline
\cite{TiaLi18} & $\SI{20}{}/\SI{2000}{}$ & \SI{2}{} & \SI{1.7}{} & \SI{0.39}{} & \SI{1.65}{} \\ \hline
\cite{HsuWan11} & $\SI{10}{}/\SI{2000}{}$ & \SI{1.7}{} & \SI{1.4}{} & \SI{0.12}{} & \SI{1.57}{} \\

\end{tabular} \label{tab:rc_exp}
\end{center}
\end{table}

In general, the contact resistivity $R_{\rm{C,1}}\cdot w_{\rm{g}}$ obtained via YFM$_1$ of the 2D FETs studied here is larger than the values obtained with YFM$_2$ using $R_{\rm{C,2}}$ of the same device. The latter are closer to reference values of \SI{0.2}{\kilo\ohm\cdot\micro\meter} \cite{WuZou16}, \SI{1.2}{\kilo\ohm\cdot\micro\meter} \cite{HsuWan11} and \SI{1.4}{\kilo\ohm\cdot\micro\meter} \cite{YarHar19} obtained with physical-based analytical \cite{WuZou16}, \cite{HsuWan11} and compact \cite{YarHar19} models, respectively, of 2D FETs describing the corresponding experimental data. YFM$_2$ also yields more similar values than YFM$_1$ results to the ones obtained via a polynomial fit of the experimental $R_{\rm{tot}}$ \cite{RoyTos14} used for the $R_{\rm{C}}$ extraction of the shorter MoS$_2$FETs: \SI{7}{\kilo\ohm\cdot\micro\meter} \cite{SanPar17}, and \SI{5}{\kilo\ohm\cdot\micro\meter} \cite{SanGho15}. The upper limit reference value of \SI{0.7}{\kilo\ohm\cdot\micro\meter} given by a TLM characterization for the shortest BPFET \cite{LiYu19} at a bias close to the device saturation regime has been accomplished by YFM$_2$ only. 

\begin{figure} [!hbtp]
\centering
\includegraphics[height=0.247\textwidth]{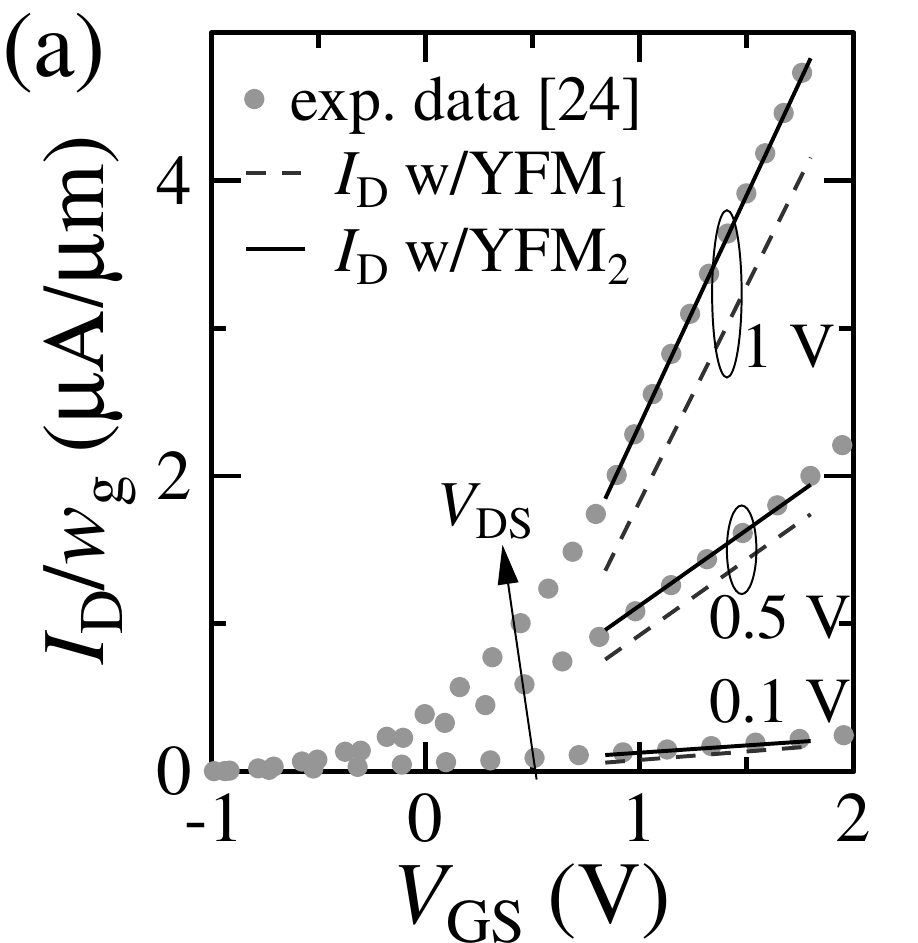}
\includegraphics[height=0.247\textwidth]{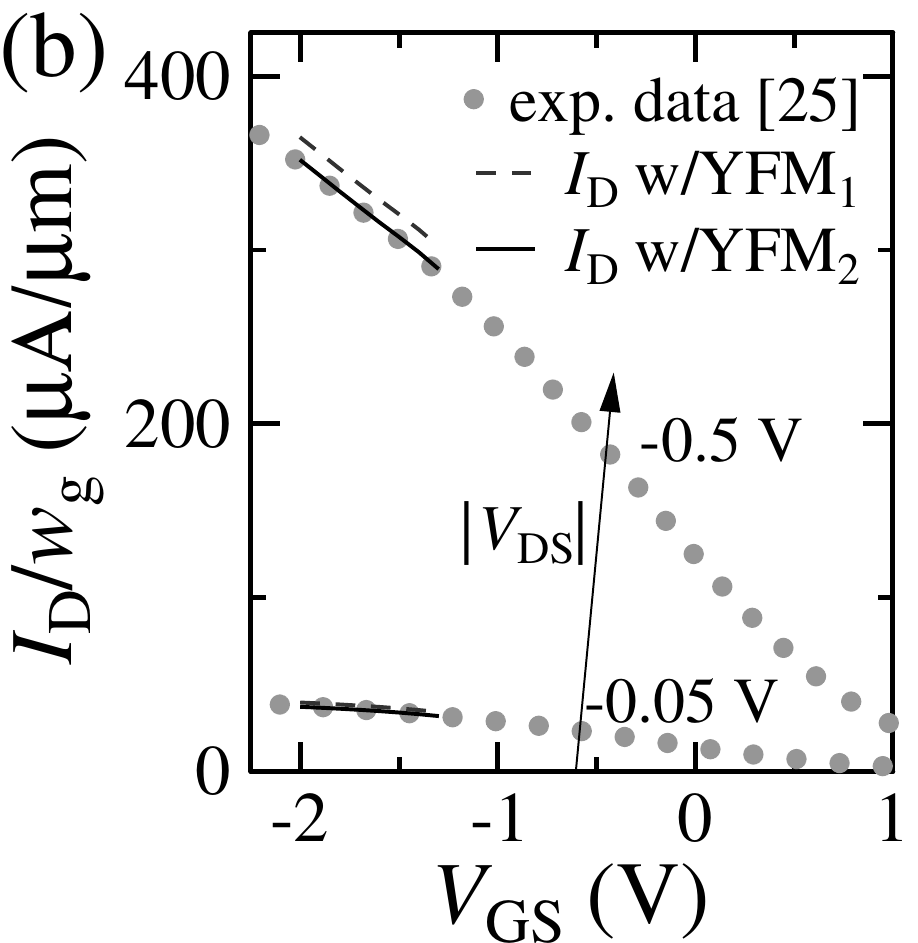} 
\includegraphics[height=0.247\textwidth]{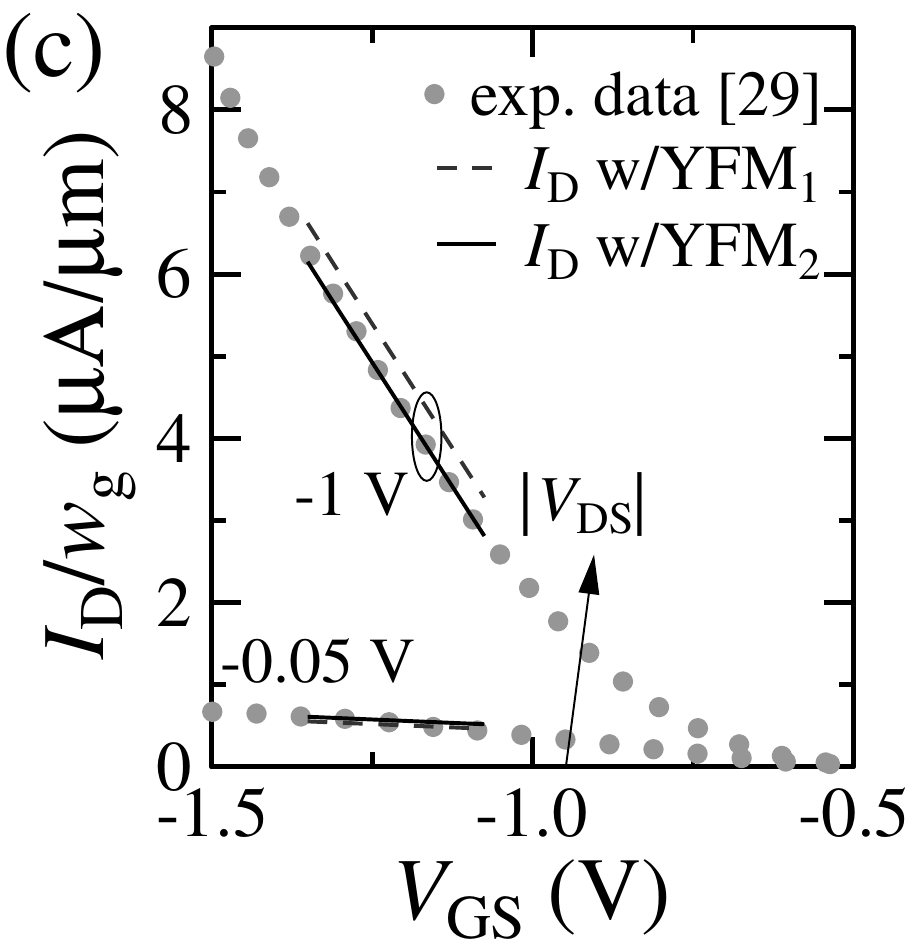}
\includegraphics[height=0.247\textwidth]{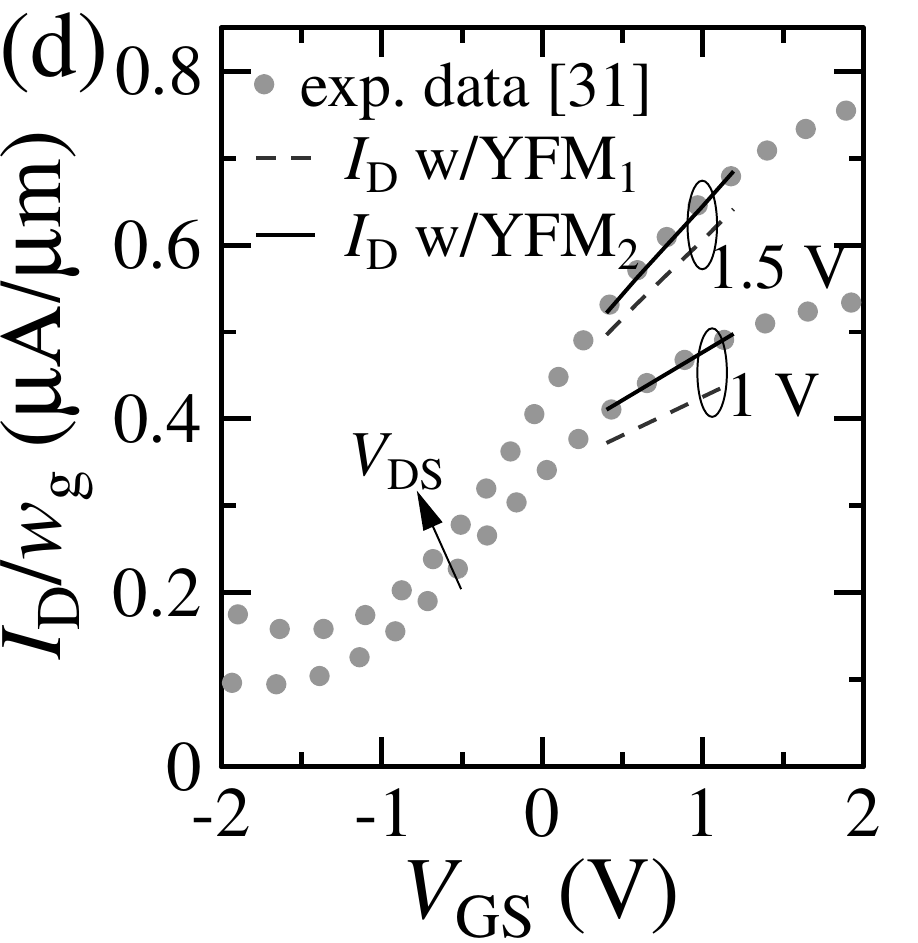} \\
\caption{Transfer characteristics of (a) a \SI{400}{\nano\meter}-long MoS$_2$FET \cite{IllBan19}, (b) a \SI{100}{\nano\meter}-long BPFET \cite{LiYu19}, (c) a \SI{9.4}{\micro\meter}-long WeS$_2$FET \cite{FanChu12} and (d) a \SI{2}{\micro\meter}-long GFET \cite{TiaLi18}. Markers are experimental data and lines represent Eq. (\ref{eq:Id_yfm}) with $\theta_{\rm{ch}}$ (solid lines) and without $\theta_{\rm{ch}}$ (dashed lines).}\label{fig:IdVg}
\end{figure}

The YFM$_1$-related condition of $\theta_{\rm{ch}} \ll \theta_{\rm{C}}$ \cite{ChaZhu14}, \cite{NaShi14} can be only considered for some of the devices studied here \cite{YarHar19}, \cite{HeiPal20}, \cite{WuZou16}, \cite{HsuWan11}. For these devices, the ratio between $R_{\rm{C,1}}$ to $R_{\rm{C,2}}$  can be between a factor of \SI{1.15}{} and \SI{2.84}{}. The overestimation of $R_{\rm{C,1}}$ with YFM$_1$ has been claimed in the literature \cite{ChaYan13}, \cite{ChaZhu14}, however, it has not been quantified before in comparison to contact resistance values considering $\theta_{\rm{ch}}$, i.e., $R_{\rm{C,2}}$. In order to further demonstrate the accuracy of each method, Fig. \ref{fig:IdVg} shows the transfer characteristics of some of the devices under study where results of Eq. (\ref{eq:Id_yfm}) calculated using the extracted parameters obtained with the different YFMs have been included within the bias region used for the extractions.


%



A factor of $\theta_{\rm{ch}}/\left[\beta (V_{\rm{GS}} - V_{\rm{th}} - \Delta) \right]$ missing in the device total resistance $R_{\rm{tot}}$ obtained with the YFM$_1$-related simplification of Eq. (\ref{eq:Id_yfm}) misleads the extracted $R_{\rm{C,1}}$ values (see Table \ref{tab:rc_exp}), i.e., the description of $I_{\rm{D}}$ is mislead with the most common extraction approach used in 2D FETs \cite{ChaYan13}-\cite{UrbLup20}. On the contrary, YFM$_2$ results describe well the experimental data due to a more complete model considered for the extraction. 
%

\begin{figure} [!hbtp]
\centering
\includegraphics[height=0.247\textwidth]{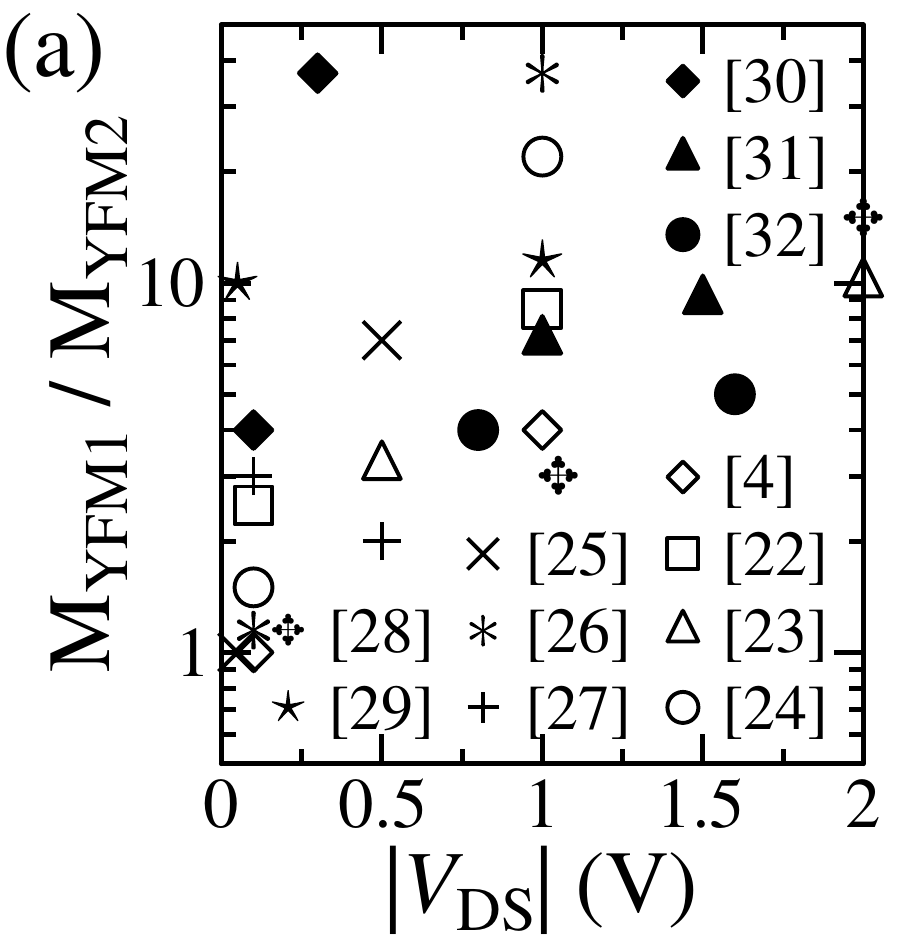}
\includegraphics[height=0.247\textwidth]{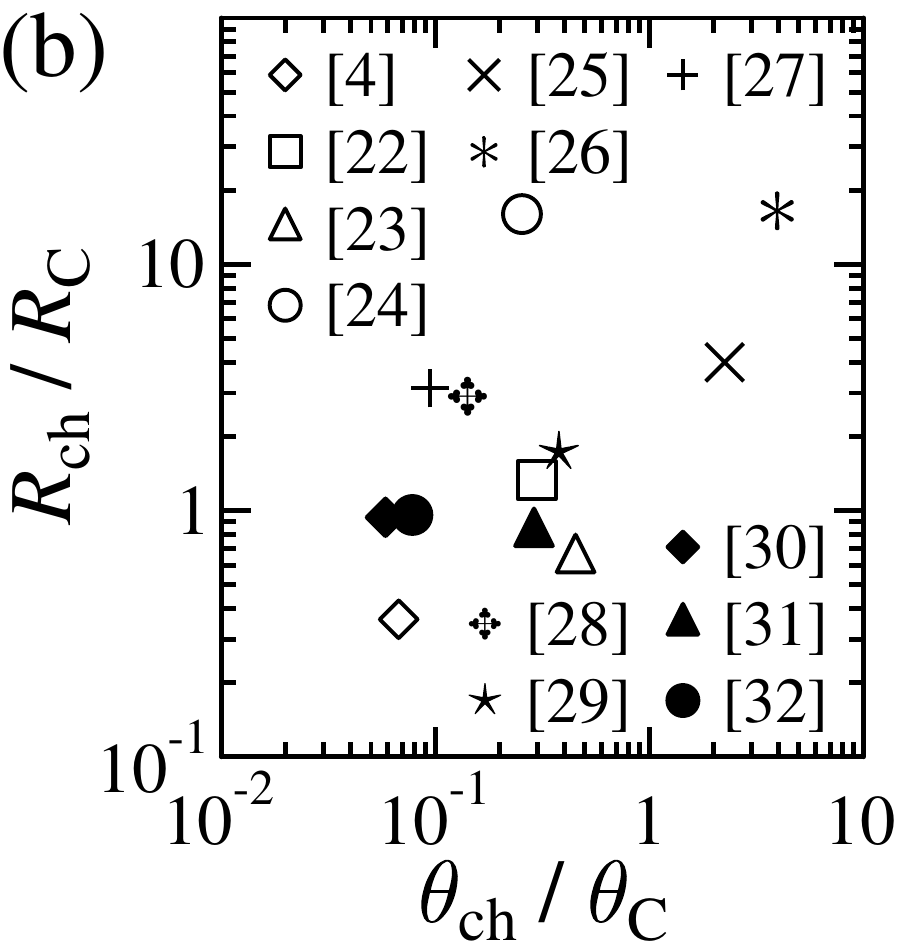} \\
\caption{Accuracy and extracted parameters of 2D FET technologies. (a) Relation of the mean absolute error over $\vert V_{\rm{DS}} \vert$ of the extraction methodologies and (b) ratios of channel to contact parameters of MoS$_2$FETs \cite{ChaYan13}, \cite{SanPar17}-\cite{IllBan19}, BPFETs \cite{LiYu19}-\cite{YarHar19}, WeS$_2$FETs \cite{HeiPal20}, \cite{FanChu12} and GFETs \cite{WuZou16}-\cite{HsuWan11}.}\label{fig:param}
\end{figure}

The mean absolute error (M) of Eq. (\ref{eq:Id_yfm}) with the parameters extracted with YFM$_1$ and YFM$_2$ with respect to the experimental data is of less than 3\% at the lowest $\vert V_{\rm{DS}}\vert$ in both cases. However a better description of the experimental data is obtained with YFM$_2$ at higher fields as shown in Fig. \ref{fig:param}(a) by the relation of M obtained with YFM$_1$ with respect to YFM$_2$ for the devices under study. The higher the $\vert V_{\rm{DS}}\vert$ the larger the error of YFM$_1$ in comparison to YFM$_{2}$. M$_{\rm{YFM2}}$ is of $\sim$1\% in all 2D FETs at the highest $\vert V_{\rm{DS}}\vert$.



The contribution of the channel properties and contact properties to the total devices performance ($R_{\rm{tot}}=R_{\rm{ch}}+R_{\rm{C}}$ and $\theta$) has been evaluated via the ratio of resistances $R_{\rm{ch}}/R_{\rm{C}}$ and mobility degradation factors $\theta_{\rm{ch}}/\theta_{\rm{C}}$ as shown in Fig. \ref{fig:param}(b). All parameters have been extracted with YFM$_2$ at the lowest available $\vert V_{\rm{DS}} \vert$. The overall static performance of most of the devices with $\theta_{\rm{C}} \ll \theta_{\rm{ch}}$ is either controlled by the contacts \cite{ChaYan13}, \cite{SanGho15}, or by both channel and contact phenomena \cite{SanPar17}, \cite{FanChu12}-\cite{HsuWan11} i.e., the accuracy of $R_{\rm{C}}$ values is relevant to understand and describe the device transport. For the devices where $\theta_{\rm{ch}}$ is more relevant than $\theta_{\rm{C}}$ \cite{LiYu19}, \cite{YanCha17}, the channel resistance is more critical for their performance. In contrast to YFM$_2$ where all these parameters can be properly characterized, YFM$_1$ yields innacurate values ($R_{\rm{C,1}}$ and $R_{\rm{ch,1}}$) and does not provide information on $\theta_{\rm{ch}}$. Hence, for modeling purposes and a precise technology evaluation, YFM$_2$ is strongly suggested over YFM$_1$.

%
%


%
%
%
%
%


\section{Conclusion}

Two different $Y$-function-based methods for the extraction of contact resistance values in 2D transistors have been described and analyzed and their results have been discussed. YFM$_1$, the most common methodology used for 2DFETs in which the intrinsic mobility degradation factor has been neglected, overestimates the contact resistance in comparison to results obtained with YFM$_2$ which considers $\theta_{\rm{ch}}$ in the underlying transport equation used for the extraction. In contrast to YFM$_1$, results of YFM$_2$ are comparable to other physics-based and experimental approaches and they are better suited to describe the experimental $I_{\rm{D}}$ within the bias region where the method has been applied. Furthermore, the factor $\theta_{\rm{ch}}$ of 2D FETs has been extracted with YFM$_2$ and it has been shown to be relevant for a proper description of $R_{\rm{C}}$ and $R_{\rm{ch}}$. YFM$_2$ has been revealed as a more rigorous and accurate methodology for the characterization of contacts of 2D FET technologies than the one usually used for these emerging transistors.


\begin{thebibliography}{}

\bibitem{Ghi88} Ghibaudo G, 'New method for the extraction of MOSFET parameters', \textit{Electronics Letters}, 1988, \textbf{24}, (9), pp. 543-545.

\bibitem{Anc10} M. G. Ancona, 'Electron transport in graphene from a diffusion-drift perspective', \emph{IEEE Transactions on Electron Devices}, 2010, \textbf{57} (3), pp. 681-689.

\bibitem{MarBad18} E. G. Marin, S. J. Bader, D. Jena, 'A New Holistic Model of 2-D Semiconductor FETs', \emph{IEEE Transactions on Electron Devices}, 2018, \textbf{65} (3), pp. 1239-1245.

\bibitem{ChaYan13} H.-Y. Chang, S. Yang, J. Lee, L. Tao, W.-S. Hwang, D. Jena, N. Lu, D. Akinwande, 'High-Performance, Highly Bendable MoS2 Transistors with High-K Dielectrics for Flexible Low-Power Systems', \emph{ACS Nano}, 2013, \textbf{7}, (6), pp. 5446-5552.

\bibitem{ChaZhu14} H.-Y. Chang, W. Zhu, D. Akinwande, 'On the mobility and contact resistance evaluation for transistors based on MoS2 or two-dimensional semiconducting atomic crystals', \emph{Applied Physics Letters}, 2014, \textbf{104}, p. 113504.

\bibitem{NaShi14} J. Na, M. Shin, M.-K. Joo, J. Huh, Y. J. Kim, H. J. Choi, J. H. Shim, G.-T. Kim, 'Separation of interlayer resistance in multilayer MoS2 field-effect transistors', \emph{Applied Physics Letters}, 2014, \textbf{104}, p. 233502.

\bibitem{BhaGan16} S. Bhattacharjee, K. L. Ganapathi, D. N. Nath, N. Bhat, 'Intrinsic Limit for Contact Resistance in Exfoliated Multilayered MoS2 FET', \emph{IEEE Electron Device Letters}, vol. 37, no. 1, pp. 119-122, Jan. 2016.

\bibitem{KimPar19} H. Kim, H. Park, G. Lee, J. Kim, 'Intimate Ohmic contact to two-dimensional WSe2 via thermal alloying', \emph{Nanotechnology}, 2019, \textbf{30}, p. 415302.

\bibitem{ParSon18} H. Park, J. Son, J. Kim, 'Reducing the contact and channel resistances of black phosphorus via low-temperature vacuum annealing', \emph{Journal of Materials Chemistry C}, 2018, \textbf{6}, pp. 1567-1572.

\bibitem{ZhuPar16} W. Zhu, S. Park, M. N. Yogeesh, K. M. McNicholas, S. R. Bank, D. Akinwande, 'Black Phosphorus Flexible Thin Film Transistors at
Gighertz Frequencies', \emph{Nano Letters}, 2016, \textbf{16}, (4), pp. 2301-2306.

\bibitem{UrbLup20} F. Urban, G. Lupina, A. Grillo, N. Martucciello, A. Di Bartolomeo, 'Contact resistance and mobility in back-gate graphene transistors', \textit{Nano Express}, 2020, \textbf{1}, p. 010001.

\bibitem{PacFei20} A. Pacheco-Sanchez, P. C. Feijoo, D. Jimenez, 'Contact resistance extraction of graphene FET technologies based on individual device characterization', (accepted) \emph{Solid-State Electronics}, 2020. Available online: https://arxiv.org/abs/2005.13926

\bibitem{DirVen20} F. Driussi, S. Venica, A. Gahoi, S. Kataria, M. C. Lemme, P. Palestri, 'Dependability assessment of Transfer Length Method to extract the metal–graphene contact resistance', \emph{IEEE Transactions on Semiconductor Manufacturing}, 2020, \textbf{33} (2), pp. 210-215.

\bibitem{PacCla16} A. Pacheco-Sanchez, M. Claus, S. Mothes, M. Schröter, 'Contact resistance extraction methods for short- and long-channel carbon nanotube field-effect transistors', \emph{Solid-State Electronics}, 2016, \textbf{125}, pp. 161-166.

\bibitem{Sch06} D. Schroder, 'Semiconductor material and device characterization', New Jersey, USA, Wiley-IEEE Press, 2016.

\bibitem{MerBor72} G. Merckel, J. Borel, N. Z. Cupcea, 'An accurate large-signal MOS transistor model for use in computer-aided design', \emph{IEEE Transactions on Electron Devices}, 1972, \textbf{19}, (5), pp. 681-690.

\bibitem{HaoCab85} C. Hao, B. Cabon-Till, S. Cristoloveanu, G. Ghibaudo, 'Experimental determination of short-channel MOSFET parameters', \emph{Solid-State Electronics}, 1985, \textbf{28}, (10), pp. 1025-1030.

\bibitem{ZhuPer09} W. Zhu, V. Perebeinos, M. Freitag, P. Avouris, 'Carrier scattering, mobilities, and electrostatic potential in monolayer, bilayer, and trilayer graphene', \emph{Physical Review B}, 2009, \textbf{80}, p. 235402.

\bibitem{WuLi16} D. Wu, X. Li, L. Luan, X. Wu, W. Li, M. N. Yogeesh, R. Ghosh, Z. Chu, D. Akinwande, Q. Niu, K. Lai, 'Uncovering edge states and electrical inhomogeneity in MoS2 field-effect transistors', \emph{Proceedings of the National Academy of Sciences of the United States of America}, 2016, \textbf{80}, (31), pp. 8583-8588.

\bibitem{KarChe14} A. Karsenty, A. Chelly, 'Application, modeling and limitations of Y-function based methods for massive series resistance in nanoscale SOI MOSFETs', \emph{Solid-State Electronics}, 2014, \textbf{92}, pp. 12–19.

\bibitem{PacCla20} A. Pacheco-Sanchez, M. Claus, 'Bias-Dependent Contact Resistance Characterization of Carbon Nanotube FETs', \emph{IEEE Transactions on Nanotechnology}, 2020, \textbf{19}, pp. 47-51.

\bibitem{SanPar17} A. Sanne, S. Park, R. Ghosh, M. N. Yogeesh, C. Liu, L. Mathew, R. Rao, D. Akinwande, S. K. Banerjee, 'Embedded gate CVD MoS2 microwave FETs', \emph{npj 2D Materials and Applications}, 2017, \textbf{1}, (26), pp. 1-6.

\bibitem{SanGho15} A. Sanne, R. Ghosh, A. Rai, M. N. Yogeesh, S. H. Shin,
A. Sharma, K. Jarvis, L. Mathew, R. Rao, D. Akinwande, S. Banerjee, 'Radio Frequency Transistors and Circuits Based on CVD MoS2', \emph{Nano Letters}, 2015, \textbf{15}, pp. 5039-5045.

\bibitem{IllBan19} Y. Y. Illarionov, A. G. Banshchikov, D. K. Polyushkin, S. Wachter, T. Knobloch, M. Thesberg, M. I. Vexler, M. Waltl, M. Lanza,
N. S. Sokolov, T. Mueller, T. Grasser, 'Reliability of scalable MoS$_2$ FETs with 2 nm crystalline CaF$_2$ insulators', \emph{2D Materials}, 2019, \textbf{6}, p. 045004.

\bibitem{LiYu19} X. Li, Z. Yu, X. Xiong, T. Li, T. Gao, R. Wang, R. Huang, Y. Wu, 'High-speed black phosphorus field-effect transistors approaching ballistic limit', \emph{Science Advances}, 2019, \textbf{5}, (6), p. eaau3194.

\bibitem{YanCha17} L. Yang, A. Charnas, G. Qiu, Y.-M. Lin, C.-C. Lu, W. Tsai, Q. Paduano, M. Snure, P. D. Ye, 'How Important Is the Metal-Semiconductor Contact for Schottky Barrier Transistors: A Case Study on Few-Layer Black Phosphorus?', \emph{ACS Omega}, 2017, \textbf{2}, pp. 4173-4179.

\bibitem{YarHar19} E. Yarmoghaddam, N. Haratipour, S. J. Koester, S. Rakheja, 'A Physics-Based Compact Model for Ultrathin Black Phosphorus FETs—Part II: Model Validation Against Numerical and Experimental Data', \emph{IEEE Transactions on Electron Devices}, 2020, \textbf{67}, (1), pp. 397-405.

\bibitem{HeiPal20} M. A. Heidarlou, P. Paletti, B. Jariwala, J. A. Robinson, S. K. Fullerton-Shirey, A. C. Seabaugh, 'Batch-Fabricated WSe2-on-Sapphire Field-Effect Transistors Grown by Chemical Vapor Deposition', \emph{IEEE Transactions on Electron Devices}, 2020, \textbf{67}, (4), pp. 1839-1844

\bibitem{FanChu12} H. Fang, S. Chuang, T. C. Chang, K. Takei, T. Takahashi, A. Javey, 'High-Performance Single Layered WSe2 p-FETs with Chemically Doped Contacts', \emph{Nano Letters}, 2012, \textbf{12}, pp. 3788-3792.

\bibitem{WuZou16} Y. Wu, X. Zou, M. Sun, Z. Cao, X. Wang, S. Huo, J. Zhou, Y. Yang, X. Yu, Y. Kong, G. Yu, L. Liao, T. Chen, '200 GHz Maximum Oscillation Frequency in CVD Graphene Radio Frequency Transistors', \emph{ACS Applied Materials \& Interfaces}, 2016, \textbf{8}, (39), pp. 25645-25649.

\bibitem{TiaLi18} M. Tian, X. Li, T. Li, Q. Gao, X. Xiong, Q. Hu, M. Wang, X. Wang, Y. Wu, 'High-Performance CVD Bernal-Stacked Bilayer Graphene Transistors for Amplifying and Mixing Signals at High Frequencies', \emph{ACS Applied Materials \& Interfaces}, 2018, \textbf{10}, (24), pp. 20219-20224.

\bibitem{HsuWan11} A. Hsu, H. Wang, K. K. Kim, J. Kong, T. Palacios, 'Impact of graphene interface quality on contact resistance and RF device performance', \emph{IEEE Electron Device Letters}, 2011, \textbf{32}, (8), pp. 1008-1010.

\bibitem{RoyTos14} T. Roy, M. Tosun, J. S. Kang, A. B. Sachid, S. B. Desai,
M. Hettick, C. C. Hu, A. Javey, 'Field-Effect Transistors Built from All Two-Dimensional Material Components', \emph{ACS Nano}, 2014, \textbf{8}, (6), pp. 6529-6564.

\end{thebibliography}
\end{document}